\documentclass[reprint,aps,twocolumn,showpacs,amsmath,amssymb]{revtex4-1}
\usepackage{graphicx,bm,float}
\def\cond{{\rm c}}
\def\val{{\rm v}}
\def\kvec{{\bf k}}
\def\kpoint{{\rm K}}
\def\kprime{{\rm K^{\prime}}}

\def\qvec{{\bf q}}
\def\kpoint{{\rm K}}
\def\kprime{{\rm K^{\prime}}}
\def\elaser{E_{\rm L}}

\def\efermi{E_{\rm F}}
\def\unitev{\ {\rm eV}}
\def\unitcm{\ \rm cm^{-1}}
\def\G{\rm G}
\def\Gprime{\rm G^\prime}
\def\Gstar{\rm G^*}
\def\iTAiTO{\rm iTA+iTO}

\begin{document}

\title{Fermi energy dependence of first- and second-order Raman
  spectra in graphene: Kohn anomaly and quantum interference
  effect}

\author{Eddwi~H.~Hasdeo$^1$} \email [Electronic Address: ]
{hasdeo@flex.phys.tohoku.ac.jp} 

\author{Ahmad~R.~T.~Nugraha$^1$}
\author{Mildred~S.~Dresselhaus$^{2,3}$}
\author{Riichiro Saito$^1$}

\affiliation{$^1$Department of Physics, Tohoku University, Sendai
  980-8578, Japan\\ $^2$Department of Electrical Engineering,
  Massachusetts Institute of Technology, Cambridge, MA 02139-4307,
  USA\\ $^3$Department of Physics, Massachusetts Institute of
  Technology, Cambridge, MA 02139-4307, USA}

\date{\today}

\begin{abstract}
  Intensity of the first- and the second-order Raman spectra are
  calculated as a function of the Fermi energy.  We show that the Kohn
  anomaly effect, i.e., phonon frequency renormalization, in the
  first-order Raman spectra originates from the phonon renormalization
  by the interband electron-hole excitation, whereas in the
  second-order Raman spectra, a competition between the interband and
  intraband electron-hole excitations takes place.  By this
  calculation, we confirm the presence of different dispersive
  behaviors of the Raman peak frequency as a function of the Fermi energy
  for the first- and the second-order Raman spectra, as observed in
  experiments.  Moreover, the calculated results of the Raman intensity
  sensitively depend on the Fermi energy for both the first- and the
  second-order Raman spectra.  These results thus also show the
  importance of quantum interference effect phenomena.
\end{abstract}

\pacs{78.67.Ch, 73.22.-f, 42.65.Dr, 03.65.Nk}
\maketitle

\section{Introduction}

Applying an electric gate voltage to graphene provides exotic tuning
of the electronic, vibrational, and optical properties of graphene
samples~\cite{ zhang08a, wang08, ukhtary15}.  Since the beginning of
graphene's discovery, electronic gating has played an important role
in elucidating the room temperature quantum Hall
effect~\cite{novoselov04, novoselov07, zhang05}, the Klein
tunneling~\cite{katsnelson06, beenakker08, popovici12}, and many body
coupling effects~\cite{sengupta08, jiang07}.  Similar gating
techniques are extensively applied not only to monolayer, but also to
multilayer graphene to obtain tunable transport~\cite{kechedzhi12}, a
tunable band gap~\cite{craciun09, zhang09}, p-n
junctions~\cite{cheianov07}, and photodetectors~\cite{xia09}.  All of
these exciting phenomena could be observed due to the ability of
tuning graphene's Fermi energy $\efermi$ through the applied gate
voltage.

A combination of electronic gating and inelastic scattering of light,
known as the gate modulated Raman spectroscopy~\cite{saito13}, opens
up a new possibility to understand more thouroughly the interplay of
the electron, photon and phonon exitations in graphene.  Several phenomena
have been probed by gate modulated Raman spectroscopy in graphene,
such as the Kohn anomaly (KA) effect or the phonon frequency
renormalization~\cite{kohn59, piscanec04, lazzeri06a, das08,
  araujo12}, the quantum interference effect~\cite{chen11, liu13},
electron-electron interaction~\cite{zhou08}, and the Fano resonance in
the Raman spectra of graphene~\cite{yoon13, hasdeo14}.  Studying gate
modulated Raman spectroscopy in graphene has also enriched our
knowledge of phonon spectra characterization~\cite{mafra12},
experimental evaluation of electron-phonon coupling~\cite{yan07}, and
various edge characterization effects~\cite{sasaki10, ren10}.

Some theoretical works have been previously performed to understand
the Kohn anomaly (KA) effect for the first-order Raman (G band)
spectra with a Raman shift of~$\sim$$1600\unitcm$ in graphene, such as
those by Ando and the Mauri's groups~\cite{ando06a,piscanec04,
  lazzeri06a}.  In the KA process, phonon renormalization occurs
through the excitation of an electron-hole pair by the electron-phonon
interaction.  As a result, the phonon energy is modified and the
phonon lifetime becomes shorter.  The previous theories mention that
the phonon frequency shows a logarithmic singularity at $T=0$~K when
the absolute value of the Fermi energy $\efermi$ matches half of the
phonon energy $|\efermi| = \hbar\omega_{\rm G}/2$.  For $|\efermi| >
\hbar\omega_{\rm G}/2$, the frequency shift is linearly proportional
to $|\efermi|$.  These predictions were already confirmed by Raman
measurements~\cite{ferrari07, yan07, das08, araujo12, saito13}.
Recently, additional experimental results allow us to study the KA
effect in the second-order Raman spectra, also.
%Focus on G' G* and G

In contrast to the first-order Raman spectra that consist of only a
single $\qvec=0$ value of the phonon momentum, the second-order Raman
spectra deals with the whole range of phonon momenta in the Brillouin
zone satisfying the double resonance Raman
condition~\cite{thomsen00}. Raman spectral features such as the
$\Gprime$ or 2D band ($\sim 2600\unitcm$) and the $\Gstar$ or $\rm
D+D^{\prime\prime}$ band ($\sim 2400\unitcm$) are observed as
the second-order Raman spectra for $\qvec\approx\kpoint$. The nonzero
momentum phonon leads to a different KA effect compared with that
for the $\qvec=0$ phonon. Araujo~\emph{et al.} and Mafra~\emph{et al.}
have shown that the frequency shift of the $\Gprime$ band as a
function of $\efermi$ is monotonically decreasing as a function of
$|\efermi|$ which is opposite to that of the G band~\cite{araujo12,
  mafra12}. The other band at $\sim 2470\unitcm$ is, however,
dispersionless as a function of $\efermi$. Yan~\emph{et al.}  show
opposite results, that the $\Gprime$ band frequency as a function of
$\efermi$ has the same trends as that of the G
band~\cite{yan07}. Further, Das~\emph{et al.}  show an asymmetric
$\Gprime$ band dispersion, i.e., a positive (negative) slope of
frequency shift at negative (positive) $\efermi$, which is
inconsistent with a symmetric dispersion shown experimentally by
Araujo~\emph{et al.}  ~\cite{das08, araujo12}. Based on the
controversies in experimental results, we present calculated results
of the second-order Raman spectra as a function of $\efermi$ from
which we understand the origin of the controvertial results.

Sasaki et al, attempted to understand why the frequency shift of the
$\Gprime$ band KA has an opposite slope when compared with that of the
$\G$ band from the viewpoint of the competition of interband and intraband
electron-hole excitation in phonon
perturbation~\cite{sasaki12a}. However, since the theory is done within
the effective mass approximation, it is not sufficient to explain the
asymmetry of the $\Gprime$ band frequency shift at positive and
negative $\efermi$. Moreover, since the Raman intensity as a function
of $\efermi$ is not calculated, different dispersion of Raman peaks as
a function of $\efermi$ cannot be explained from that theory.

Observing the change of Raman intensity as a function of $\efermi$
reveals the quantum interference effect. When $\efermi\neq 0$, some
Raman scattering paths are suppressed due to the Pauli exclusion
principle. Even with the reduced number of scattering paths, the Raman
intensity surprisingly increases at a particular value of $\efermi$
when constructive Raman phases among various scattering paths are
enhanced. Chen~\emph{et al.}  show that the G band Raman intensity
gives a maximum value when $2|\efermi|=\elaser-\hbar\omega_G/2$, where
$\elaser$ is the laser energy~\cite{chen11}. However, the theoretical
analysis in their work assumes a constant matrix element therefore
neglecting the change of the Raman phase due to the electron-phonon
matrix elements. Previous theoretical calculations show that the
electron-phonon matrix elements change sign along electronic
equi-energy lines in graphene and therefore can change the Raman
phase~\cite{jiang04,sasaki08a}. A comprehensive calculation is,
therefore, necessary to understand how the quantum interference effect
affects the first- and the second-order Raman intensity.

In this work we calculate the $\efermi$ dependece of the first-order
and the second-order Raman spectra. The calculated spectral quantities
are the Raman peak shift, spectral linewidth, and the Raman intensity
as a function of $\efermi$. The KA correction including both the phonon
frequency shift and the linewidth is modeled based on  second-order
perturbation theory. The KA of the first-order Raman spectra or
of the $\qvec=0$ phonon is calculated so as to reproduce the existing
theoretical and experimental results and to compare with the KA of
the $\qvec\neq0$ phonon.%  The second-order Raman intensity is calculated
% following the work by Venezuela~\emph{et al.}~\cite{venezuela11}.  
We now focus on the intervalley scatterings which give three prominent
peaks in the experimental spectra, namely, the $\Gprime$, the $\Gstar$, and
the $\iTAiTO$ bands, and are relevant to $\qvec\approx\kpoint$. The
$\elaser$ dependence of those Raman peak positions is compared with
experimental results in order to justify the present calculation
methods. Finally the $\efermi$ dependence of those three Raman spectra
are analized and compared to the experimental results.

\begin{figure}[t]
\includegraphics[width=8cm]{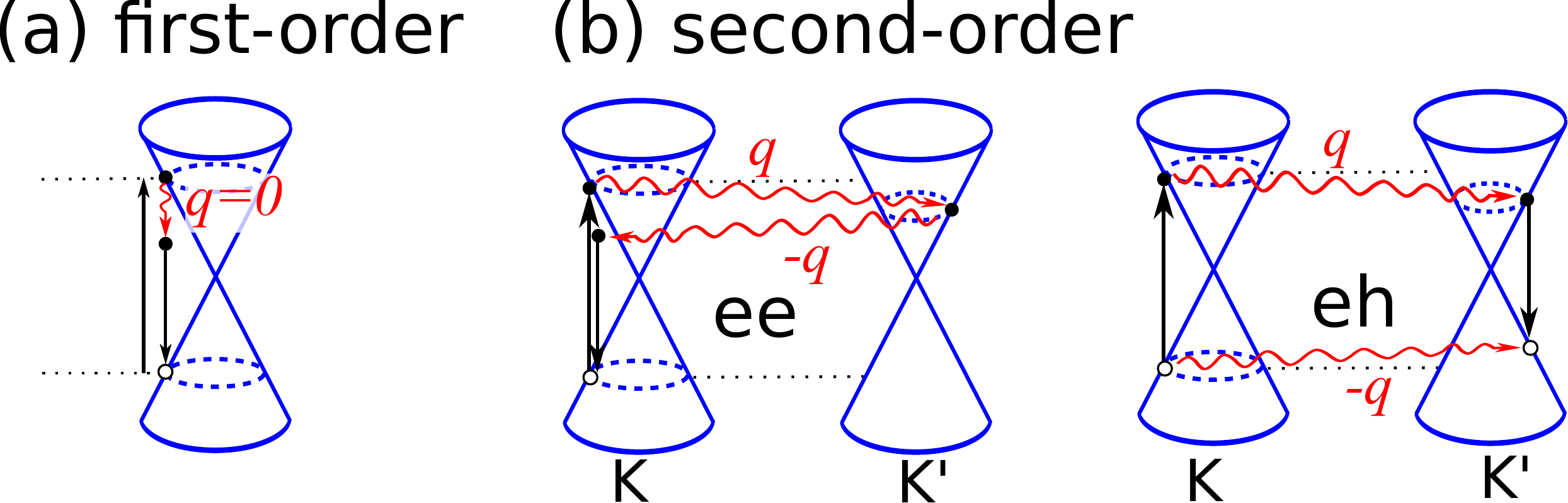}
\caption{\label{Fig1}(Color online) Schematics of (a) the first- and
  (b) the second-order Raman process. In (b), ee means two consecutive
  electron-phonon interactions while eh means electron-phonon
  interaction followed by hole-phonon interaction.}
\end{figure}

The organization of this paper is as follows.
Section~\ref{subsec:ramanintens} shows the method for calculating the
Raman intensities for the first- and second-order Raman
spectra. Section~\ref{subsec:ka} explains the method for numerically
calculating the KA effect. Section~\ref{subsec:first} presents the
calculated results of the KA effect for the $\qvec=0$ phonon and the
$\efermi$ dependence of the G band intensity.
Section~\ref{subsec:second} presents the calculation results of the KA
effect for $\qvec\neq0$, and the $\elaser$ and $\efermi$ dependences of
the second-order Raman spectra.  Finally, the conclusion is given in
Sect.~\ref{sec:conclusion}.

\section{Calculation Methods}
\label{sec:calmeth}
\subsection{Raman Intensity}
\label{subsec:ramanintens}

The first-order Raman process as shown in Fig.\ref{Fig1}(a) consists
of (1) excitation of an electron-hole pair by the electron-photon
interaction, (2) phonon emission by means of the electron-phonon
interaction, and (3) electron-hole recombination and photoemission by
the electron-photon interaction. Based on the three subprocesses,
the Raman intensity formula for the first-order Raman process is given
by
\begin{align}
  \label{eq:gband}
  I^{(1)}=& \sum_\nu \left |\sum_\kvec \frac{M_{\rm op}^{\val\cond}(\kvec)
      M_{\rm ep}^{{\rm eh}\nu}(\kvec,\kvec)
      M_{\rm op}^{\cond\val}(\kvec)
      \left[f(E_\kvec^\val)-f(E_\kvec^\cond)\right]}
    {(\elaser-E_\kvec^{\cond\val}-i\gamma)
      (\elaser-E_\kvec^{\cond\val}-\omega^\nu_0-i\gamma)}\right|^2 \nonumber\\
  &\times\delta(\elaser-\omega^\nu_0-E_{\rm s}), 
\end{align}
where $\elaser$ is the laser excitation energy, $E_{\rm s}$ is
the scattered photon energy, $E_\kvec^{\cond\val}
=E_\kvec^{\cond}-E_\kvec^{\val}$ is the electron energy difference between
the conduction ($\rm c$) and the valence ($\rm v$) bands at a wave vector
$\kvec$, $\gamma = (37.6\elaser+13.6\elaser^2)\times10^{-3}\ {\rm eV}$
is the carrier scattering rate due to the electron-phonon
interaction~\cite{venezuela11}, and $f(E)$ is the Fermi distribution
function which depends on temperature.  $M_{\rm
  op}^{\cond\val}(\kvec)$ is the electron-photon matrix element
acounting for the optical transition of an electron in a state $\kvec$
from a valence band to a conduction band, $M_{\rm ep}^{{\rm
    eh}\nu}({\bf q},{\bf p})= M_{\rm ep}^{{\rm cc}\nu}({\bf q},{\bf
  p})- M_{\rm ep}^{{\rm vv}\nu}({\bf q},{\bf p})$ is the
carrier-phonon interaction considering an electron ($\rm e$) in a
conduction band or a hole ($\rm h$) in a valence band making a
transition from a state $\bf p$ to a state $\bf q$ by emitting a
phonon with momentum ${\bf q}-{\bf p}$, mode $\nu$, and frequency
$\omega^\nu_{\qvec-{\bf p}}$.  Hereafter, $\hbar = 1$ is used, so that
$\omega^\nu_{\qvec-{\bf p}}$ has units of energy.  For the case of a
one phonon process, only zero momentum or the $\Gamma$ point phonon is
relevant. The summation over $\kvec$ in Eq.~\eqref{eq:gband} is taken
to occur in a uniform square mesh, with a mesh spacing $\Delta k=
\gamma/2v$, and $v=6.46\ \rm eV\AA$ is the slope of the electron energy
dispersion of graphene and $\Delta k^2$ is the weight of the integration. We set a cutoff
energy $E_{\rm cut}=3.5\ \rm eV$ for $E_\kvec^{\cond\val}$ so as to
reduce the number of mesh points in the Brillouin zone integration.  It is
important to note that both the numerator and denominator of
Eq.~\eqref{eq:gband} are complex numbers, thus the summation of
$\kvec$ before taking the square plays an important role in the quantum
interference effect~\cite{martin83, maultzsch04}.

\begin{figure}[t]
\includegraphics[width=8cm]{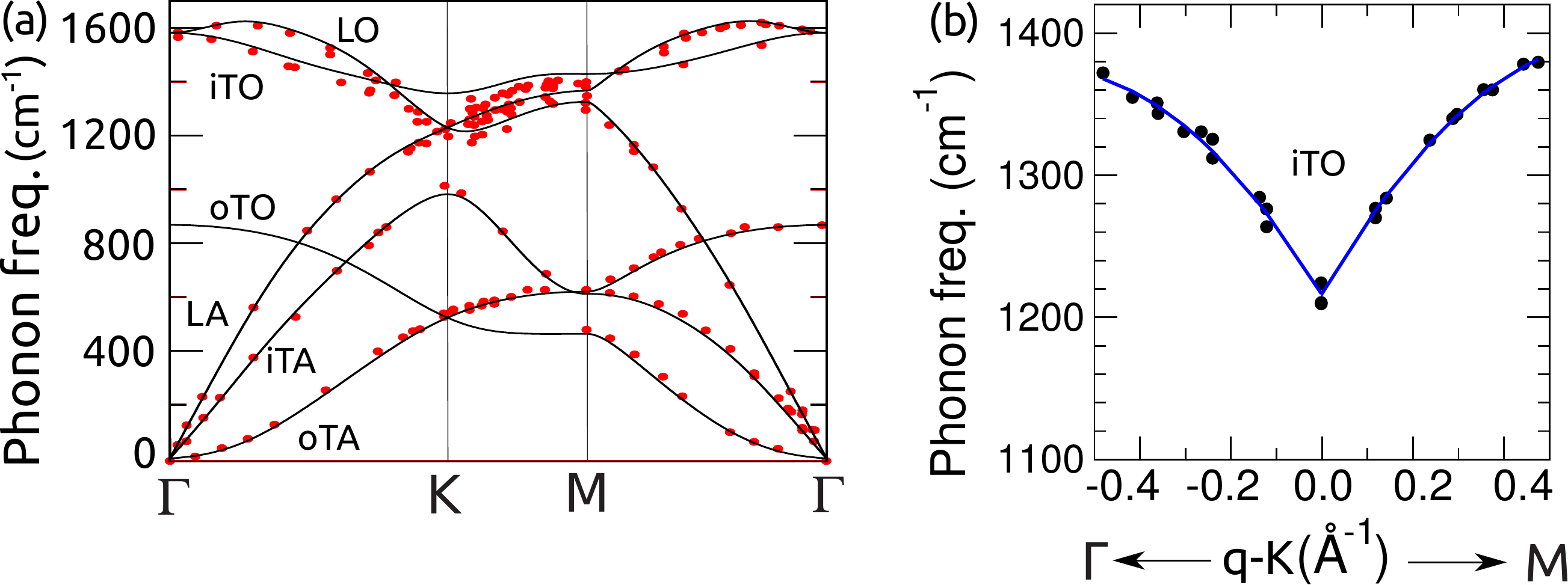}
\caption{\label{Fig-phdisp}(Color online) (a) Calculated (solid lines,
  this work) and experimental (red dots, from
  Refs.~\onlinecite{maultzsch04a, mohr07}) phonon dispersion
  relations. (b) Fitting of Eq.~\eqref{eq:wito} (blue line) to the iTO
  branch from Ref.~\onlinecite{gruneis09b} (dots) near the K point.}
\end{figure}

The electronic structure of graphene is calculated within an extended
tight binding method considering up to 20 nearest neighbors and the
four atomic orbitals ($2p_x$, $2p_y$, $2p_z$, $2s$)~\cite{popov04a,
  samsonidze04}. Calculation of the phonon dispersion relations is
performed within a force constant model with the interatomic potential
including up to 20 nearest neighbors which is fitted from a
first-principles calculation~\cite{dubay03,
  saito10}. Figure~\ref{Fig-phdisp}(a) shows the calculated results of
the phonon dispersion relations (solid lines) and the corresponding
experimental phonon dispersion relations (red dots) for comparison
from Refs.~\onlinecite{maultzsch04a, mohr07}. Because of the KA
effect, the dispersion of the in-plane tangential optic (iTO) branch
near the $\kpoint$ point is discontinuous along the $\rm \Gamma-K-M$
path which cannot be reproduced by the force constant
model~\cite{piscanec04}. We fit the iTO frequency from the
experiment~\cite{gruneis09b} and use the following fitting formula for
the Raman spectra calculation [Fig.~\ref{Fig-phdisp}(b)]:
\begin{align}
  \label{eq:wito}
  \omega^{\rm iTO}_\qvec=& \big\{-424.81 q^2+ 534.47 q + 1215.95
  \nonumber\\
  &+(6.94q^2+10.89q)\cos(3\theta)\big\}\unitcm,
\end{align}
where $\qvec$ is defined using polar coordinates $(q,\theta)$ whose
center is at the K point and $\theta$ is measured from the KM
direction. Eq.~\eqref{eq:wito} is valid only for $q\le 0.4\ {\rm
  \AA^{-1}} $, and when $q>0.4\ {\rm \AA^{-1}} $, we use the results from
the force constant model for $\omega^{\rm iTO}_\qvec$.

In the electron-photon interaction, we adopt a dipole approximation,
i.e., assuming a slowly varying electric field in space because the laser
wavelength is much greater than the inter-atomic
distance~\cite{gruneis03}. The electron-phonon interaction is calculated
using the tight binding method with the deformation potential fitted
from the GW method for the K and $\Gamma$ points~\cite{venezuela11}.

In the second-order Raman processes, phonons with modes $\nu$ and
$\mu$ and momenta $\qvec$ and $-\qvec$, respectively, are emitted
[Fig.~\ref{Fig1}(b)]. Depending on the carriers taking part in the
scattering event, the Raman intensity formula is given by:
\begin{align}
  \label{eq:twoph}
  I^{(2)} =& \sum_{\qvec\nu\mu}\big|A_{\qvec\nu\mu}^{\rm
      ee}+A_{\qvec\nu\mu}^{\rm hh}+A_{\qvec\nu\mu}^{\rm
      he}+A_{\qvec\nu\mu}^{\rm eh} \nonumber\\
  &+A_{-\qvec\mu\nu}^{\rm
      ee}+A_{-\qvec\mu\nu}^{\rm hh}+A_{-\qvec\mu\nu}^{\rm
      he}+A_{-\qvec\mu\nu}^{\rm eh}\big|^2 \nonumber\\
  &\times \delta(\elaser-\omega_\nu-\omega_\mu-E_{\rm s}),
\end{align}
where $A_{\qvec\nu\mu}^{\rm eh}$ is a Raman amplitude for each process:
(1) an electron (e), first, emits a $\nu$ phonon with momentum
$\qvec$ and, (2) a hole (h) emits the $\mu$ phonon with momentum
$-\qvec$. Here, $A_{\qvec\nu\mu}^{\rm eh}$ and $A_{-\qvec\mu\nu}^{\rm eh}$
are not equivalent to each other due to the different time order of
the two phonon emission. Next, we show examples of the Raman amplitude formula
for $A_{\qvec\nu\mu}^{\rm ee}$ and $A_{\qvec\nu\mu}^{\rm eh}$:
\begin {widetext}
  \begin{equation}
    \label{eq:ampliee}
    A_{\qvec\nu\mu}^{\rm ee}=\sum_k\frac{M_{\rm op}^{\val\cond}(\kvec)
      M_{\rm ep}^{{\rm cc} \mu}(\kvec,\kvec+\qvec)M_{\rm ep}^{{\rm cc}\nu}(\kvec+\qvec,\kvec)M_{\rm
        op}^{\cond\val}(\kvec)
      \left[f(E_\kvec^\val)-f(E_\kvec^\cond)\right]}
    {(\elaser-E_\kvec^{\cond\val}-i\gamma)
      (\elaser-E_{\kvec+\qvec}^{\cond}+E_{\kvec}^{\val}-\omega^\nu_{-\qvec}-i\gamma)
      (\elaser-E_{\kvec}^{\cond\val}-\omega^\nu_{-\qvec}-\omega^\mu_{\qvec}-i\gamma)},
  \end{equation}
  
  \begin{equation}
    \label{eq:amplieh}
    A_{\qvec\nu\mu}^{\rm eh}=-\sum_k\frac{M_{\rm op}^{\val\cond}(\kvec+\qvec)
      M_{\rm ep}^{{\rm vv}\mu}(\kvec+\qvec,\kvec)M_{\rm ep}^{{\rm
          cc}\nu}
      (\kvec+\qvec,\kvec)M_{\rm
        op}^{\cond\val}(\kvec)
      \left[f(E_\kvec^\val)-f(E_\kvec^\cond)\right]}
    {(\elaser-E_\kvec^{\cond\val}-i\gamma)
      (\elaser-E_{\kvec+\qvec}^{\cond}+E_{\kvec}^{\val}-\omega^\nu_{-\qvec}-i\gamma)
      (\elaser-E_{\kvec+\qvec}^{\cond\val}-\omega^\nu_{-\qvec}-\omega^\mu_{\qvec}-i\gamma)}.
  \end{equation}
  
\end{widetext}
The minus sign in Eq.~\eqref{eq:amplieh} corresponds to the opposite charge of
the hole from the electron in the hole-phonon matrix
elements~\cite{jiang07a}.

\begin{figure}[t!]
  \includegraphics[width=8cm]{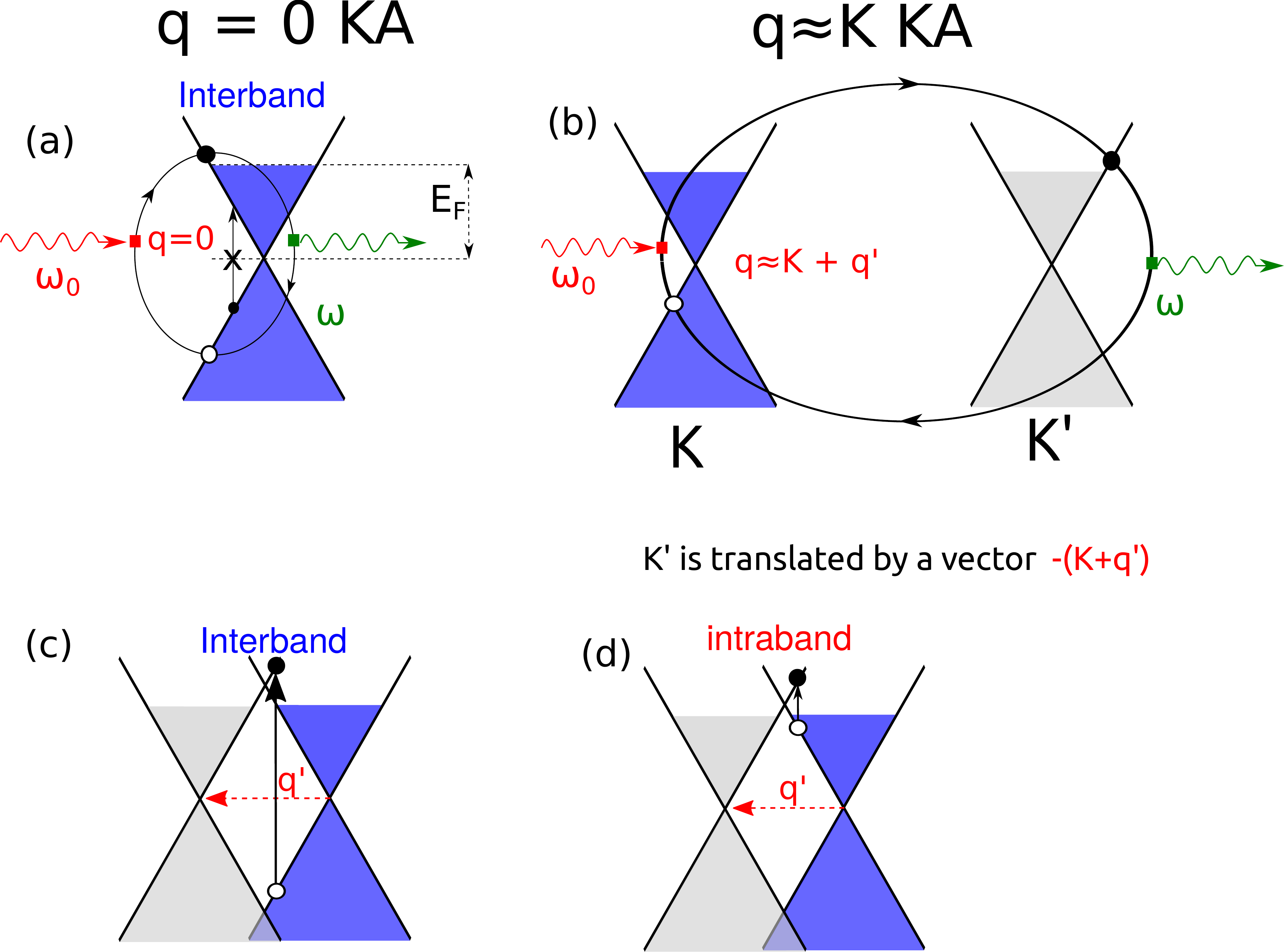}
  \caption{\label{Fig2a}(Color online) (a) A schematic of the $\qvec
    =0$ Kohn anomaly process. A phonon with zero wave vector ($\qvec
    =0$) and frequency $\omega_0$ vertically excites an electron-hole
    pair via the electron-phonon interaction. The electron-hole pair
    then recombines by emitting a phonon with frequency $\omega$.  (b)
    A schematic of the $\qvec \approx \kpoint $ Kohn anomaly
    process. An electron exists at the $\kprime$ point leaving a hole
    behind at the $\kpoint$ point with a distance in reciprocal space
    of $\qvec={\bf\kpoint}+\qvec^\prime$. If the $\kprime$ point is
    then translated by a vector $-({\bf K}+\qvec^\prime)$, we can then
    imagine a virtual vertical transition of electron and hole. When
    $\efermi \neq 0$, both interband (c) and intraband (d) transitions
    are expected.}
\end{figure}
\subsection{The Kohn Anomaly}
\label{subsec:ka}
Kohn mentions that conduction electrons are able to screen phonons in
a metal~\cite{kohn59}. This screening leads to a phonon frequency
change, given by:
\begin{equation}
\label{eq:correc}
\omega^\nu_{\qvec}=\omega^{(0),\nu}_{\qvec}+\omega^{(2),\nu}_{\qvec},
\end{equation}
where $\omega^{(0),\nu}_{\qvec}$ is the unperturbed phonon energy from
the phonon dispersion relation. Here, $\omega^{(2),\nu}_{\qvec}$ is the
correction term taken from the second-order perturbation of
the electron-phonon interaction by the excitation and recombination of an
electron-hole pair (Fig.~\ref{Fig2a}):
\begin{equation}
\label{eq:KA}
\omega^{(2),\nu}_{\qvec}= 2\sum_{s,s'}^{\cond,\val}\sum_\kvec
\frac{|M_{\rm ep}^{ss'\nu}(\kvec,\kvec+\qvec)|^2
\left[f(E_\kvec^s)-f(E_{\kvec+\qvec}^{s'})\right]}
{\omega^{(0),\nu}_{\qvec}-E_{\kvec+\qvec}^{s'}+E_\kvec^{s}+i\eta},
\end{equation}
where the prefactor $2$ in Eq.~\eqref{eq:KA} accounts for the spin
degeneracy, while the valley degeneracy is considered in the summation
over $\kvec$. The value of $\omega^{(2),\nu}_{\qvec}$ is a complex
number, in which ${\rm Re}(\omega^{(2),\nu}_{\qvec})$ [ $-{\rm
    Im}(\omega^{(2),\nu}_{\qvec})$] gives the phonon frequency shift
[phonon linewidth]. In Eq.~\eqref{eq:KA}, the contribution of the
interband (intraband) electron-hole pair appears at $s\neq s'$
($s=s'$).

In a conventional 2D electron gas, the KA effect occurs at $\qvec =
2\kvec_{\rm F}$, where $\kvec_{\rm F}$ is the Fermi wave vector. In
graphene, due to its unique linear energy bands, the KA occurs at
$\qvec \approx 0$ and $\qvec \approx \kpoint$. The schematics of the
KA process for $\qvec = 0$ and $\qvec \approx \kpoint $ are shown in
Figs.~\ref{Fig2a}(a) and (b), respectively. In the $\qvec =0$ KA, a
phonon with frequency $\omega_0$ vertically excites an electron-hole
pair via the electron-phonon interaction [Fig.~\ref{Fig2a}(a)]. The
electron-hole pair then recombines by emitting a phonon with frequency
$\omega$.  In the $\qvec \approx \kpoint$ KA, an electron exists at
the $\kprime$ point, leaving a hole behind at the $\kpoint$ point with
a distance in reciprocal space $\qvec=\kpoint+\qvec^\prime$
[Fig.~\ref{Fig2a}(b)]. If the $\kprime$ point is translated by a
vector $-({\bf K}+\qvec^\prime)$, we can imagine a virtually vertical
excitation of an electron-hole pair. When $\efermi \neq 0$, both the
interband [Fig.~\ref{Fig2a}(c)] and intraband [Fig.~\ref{Fig2a}(d)]
transitions are expected.

\section{Results and Discussion}
\label{sec:resul}
\subsection{First-order Raman spectra}
\label{subsec:first}
\begin{figure}[t]
\includegraphics[width=8cm]{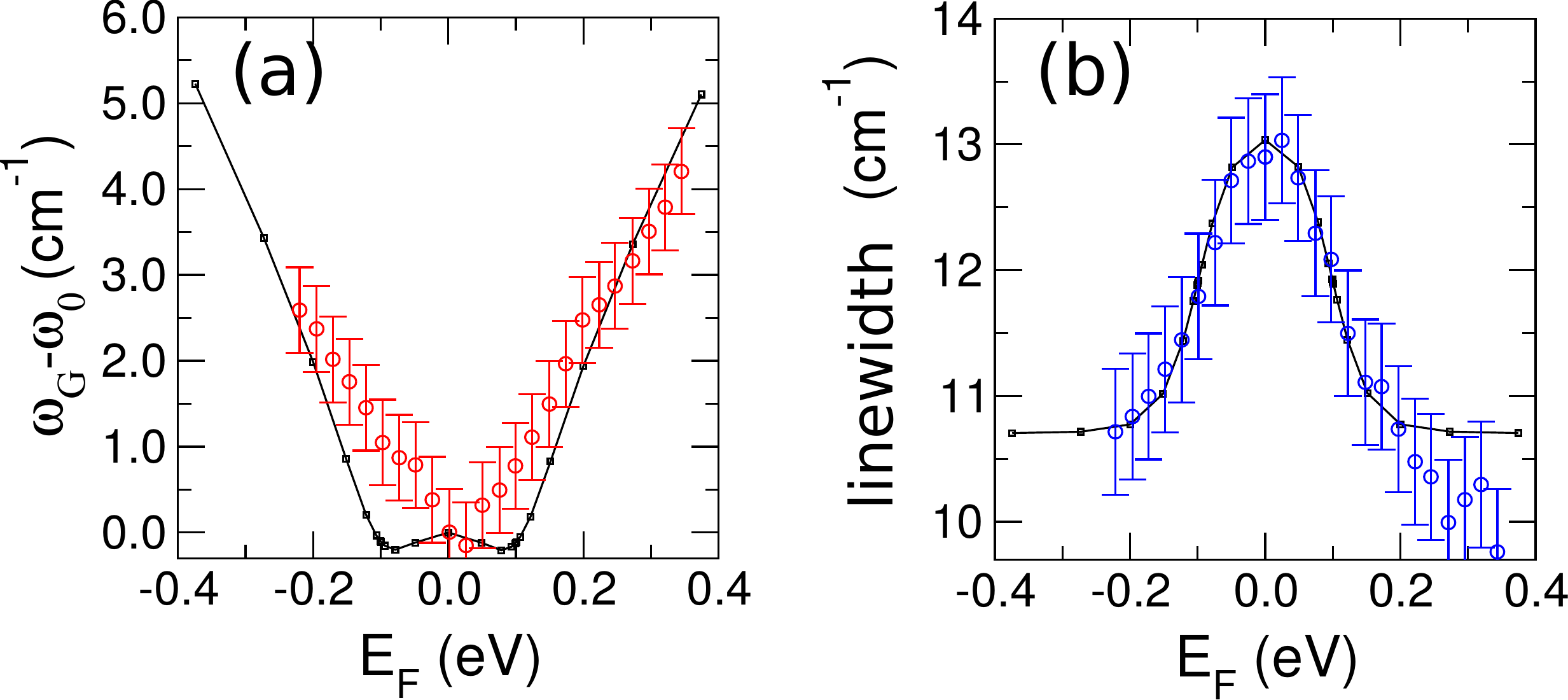}
\caption{\label{Fig2}(Color online) The calculated (dotted
  line) and experimental (open circles) results for the G band peak
  shift (a) and the G band linewidth (b) as a function of the Fermi energy,
  respectively, for $T = 300$ K.}
\end{figure}

Employing Eq.~\eqref{eq:KA} at $\qvec=0$, we can obtain the frequency
shift [Fig.~\ref{Fig2}(a)] and phonon linewidth [Fig.~\ref{Fig2}(b)]
for the G band as a function of the Fermi energy at $T =
300$~$\kpoint$. In Fig.~\ref{Fig2} we show the calculated (dotted
line) and experimental (open circles) results~\cite{araujo12} of the G
band peak shift and linewidth as a function of the Fermi energy,
respectively. The calculated results are in good agreement with the
experimental results. In Fig.~\ref{Fig2}(a), we see dips when
$2|\efermi|=\omega_0\approx 0.2\ \rm eV$ for the calculation, while
the experimental results do not show such dips. These dips are
originated from the logarithmic singularities at $T=0$~$\kpoint$ and
are related to interband resonances~\cite{ando06a, sasaki08,
  lazzeri06a}. For $2|\efermi|>\omega_0$, the G band frequency
increases linearly as a function of the Fermi energy. At $0$ K, the
phonon linewidth shows a step function
$\theta(\omega_0-2|\efermi|)$. The step function indicates that when
$2|\efermi|>\omega_0$, the phonon linewidth from the KA effect becomes
zero since no excited electron-hole pair meets the resonance condition
of Eq.~\eqref{eq:KA}. At finite $T$, on the other hand, the Fermi
distribution function becomes a smooth function and that is why we get
a smooth function of the linewidth as a function of $\efermi$. It is
noted that we add an extrinsic broadening of $10.3\unitcm$ in our
calculations in order to fit with experimental results~\cite{araujo12} in
Fig.~\ref{Fig2}(b).

\begin{figure}[t]
\includegraphics[width=8cm]{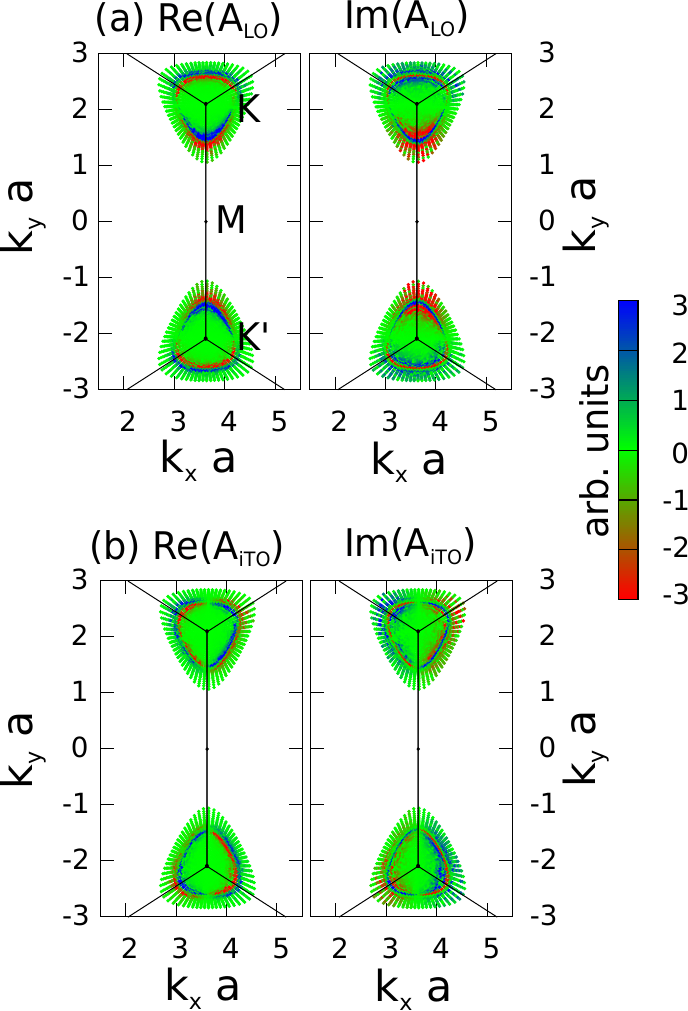}
\caption{\label{Fig3}(Color online) Calculated results of the real and
  imaginary parts of the first-order (a) LO and (b) iTO Raman amplitudes in
  a two dimensional Brillouin zone near the $\kpoint$ and $\kprime$
  points with $\elaser = 2.33$ eV.}
\end{figure}

Next, we calculate the Raman spectra of the G band using
Eq.~\eqref{eq:gband}. The G band consists of both the $\qvec = 0$
longitudinal optic (LO) and in-plane-tangential optic (iTO) modes. In
order to understand their contributions to the Raman amplitudes at
each $\kvec$ point, we plot the real and imaginary parts of the Raman
amplitude in Eq.~\eqref{eq:gband} for LO and iTO phonons in
Fig.~\ref{Fig3}(a) and (b), respectively. Here we use $\elaser = 2.33$
eV and take $E_{\rm cut}=3.5\unitev$ so as to reduce the total points
of integration for saving computational time. It will be clear that
neglecting the contributions from energies above $E_{\rm cut}$ in the
integration is reasonable since the Raman intensity is quickly
decreasing when $2|\efermi|>E_{\rm cut}$.  In Fig.~\ref{Fig3},
deformed triangles near the $\kpoint$ and the $\kprime$ points
indicate equi-energy lines that match the resonant conditions. The
lower (higher) resonant condition corresponds to the scattered
(incident) resonance when $E_\kvec^{\cond\val}=\elaser-\omega_G$
($E_\kvec^{\cond\val}=\elaser$) which is shown by a large amplitude at
the inner (outer) line. We see changes in the sign for both the real
and imaginary parts of the LO and iTO Raman amplitudes in both the
radial and azimuthal directions. The change of sign at the radial
direction is related to an opposite phase between the scattered
resonance and the incident resonance. Meanwhile, the change of sign in the
azimuthal direction is related to the sign of the electron-phonon
matrix element as reported by Jiang et al~\cite{jiang04}. The LO (iTO)
phonon has a zero matrix element at the 0 ($\pi/2$) phase.

\begin{figure}[t]
\includegraphics[width=8cm]{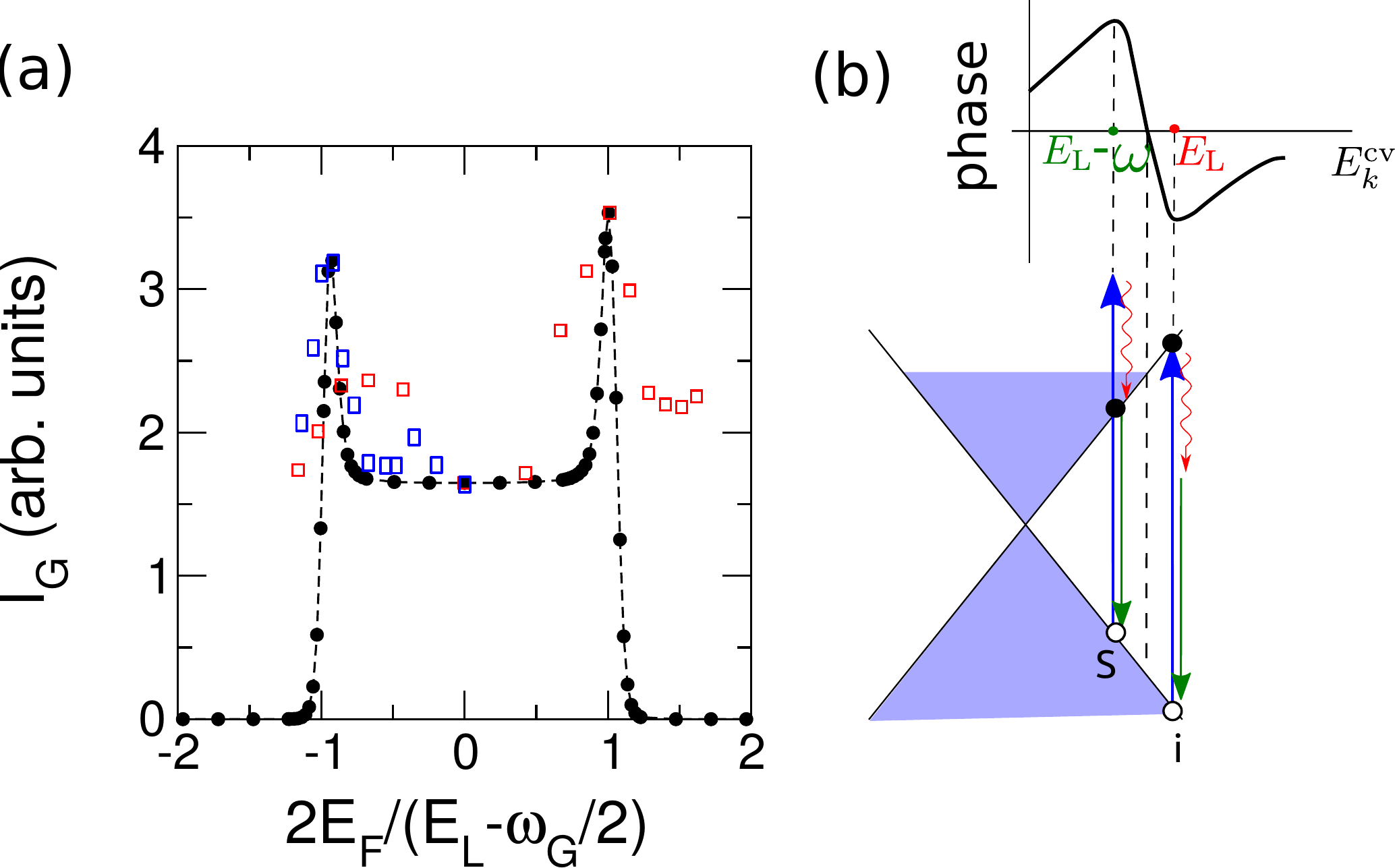}
\caption{\label{Fig4}(Color online) (a) Calculated results (black
  dots) and experimental results (blue squares from
  Ref.~\onlinecite{chen11} and red squares from
  Ref.~\onlinecite{liu13}) of the G band Raman intensity as a function
  of the reduced Fermi energy. (b) Schematic diagram showing an
  opposite phase between the incident (i) and scattered (s)
  resonances. When $2|\efermi|=\elaser-\omega_G/2$, the scattered
  resonance is suppressed, and therefore, the Raman intensity gives a
  maximum value.}
\end{figure}

The opposite phase of the scattered resonance to the incident
resonance is essential insofar as both terms give destructive
interference. Therefore, only taking the resonant term for calculating
the G band intensity is not sufficient. We need to at least consider
up to $E_\kvec^{\cond\val}\approx\elaser+\omega_G$ to get a realistic
intensity. Moreover if we plot the Raman intensity as a function of
the Fermi energy as shown in Fig.~\ref{Fig4}(a), it becomes clear that
destructive interference between the scattered resonance and the
incident resonance can be suppressed when we set the Fermi energy
close to the laser excitation energy. When
$2|\efermi|=\elaser-\omega_G/2$, the scattered resonance cannot occur
due to the Pauli blocking effect [Fig.~\ref{Fig4}(b)]. Therefore, in
Fig.~\ref{Fig4}(a) we see the largest G band intensity at
$2|\efermi|=\elaser-\omega_G/2$ as pointed out by Chen~\emph{et
  al}~\cite{chen11}. The difference of the intensity at positive and
negative $\efermi$ which comes from the electron-hole asymmetry has
been confirmed by Liu~\emph{et al}~\cite{liu13}. Anisotropy in the
azimuthal direction should give destructive interference, but the
effect is negligible.

\begin{figure*}[t]
\includegraphics[width=12cm]{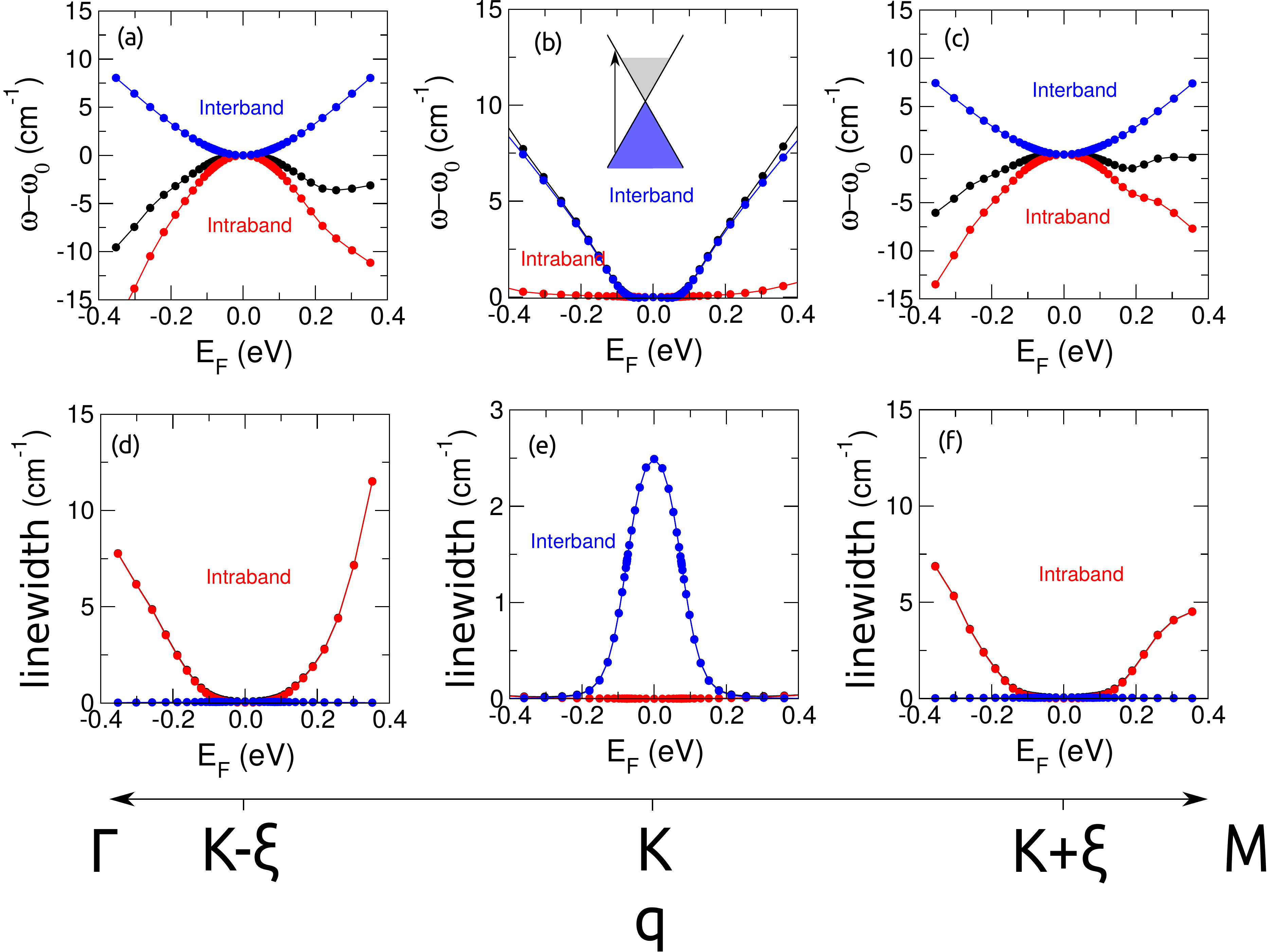}
\caption{\label{Fig5}(Color online) The iTO phonon energy shift and
  linewidth as a function of the Fermi energy $\efermi$ for (a), (d) $\qvec=
  \kpoint-\xi$; (b), (e) $\qvec= \kpoint$; and (c), (f) $\qvec=
  \kpoint+\xi$, with $\xi= 0.14\ \rm \AA^{-1}$. We use $T= 300\ \rm
  K$. }
\end{figure*}
\subsection{Second-order Raman spectra}
\label{subsec:second}
In Fig.~\ref{Fig5}, we show the calculated results of the $\qvec\neq0$
KA effect from Eq.~\eqref{eq:KA}. First, let us consider the case of
$\qvec=\kpoint$ in Figs.~\ref{Fig5}(b) and (e). If we compare
respectively Figs.~\ref{Fig5}(b) and (e) with Figs.~\ref{Fig2}(a) and
(b), both the frequency shift and phonon linewidth show the same
trends as that of $\qvec=0$ KA because both $\qvec=0$ and
$\qvec=\kpoint$ are dominated by the interband electron-hole
excitation. The reason why the interband excitation is dominant at
$\qvec = \kpoint$, is that the $\kpoint$ and $\kprime$ points coincide
upon translation of the $\kprime$ point by a vector $-\kpoint$ [
  $\qvec^\prime=0$ in Fig.~\ref{Fig2a}(c)]. Therefore, at the $\qvec =
\kpoint$ KA, only virtually vertical interband excitation, the same as
at $\qvec = 0$ KA, is possible~\cite{sasaki12}. The previous work did
not consider the interband contribution, therefore assigning the
$\qvec=\kpoint$ phonon frequency shift to be dispersionless as a function
of $\efermi$~\cite{araujo12}.

Next, if we shift by $\xi= 0.14\ \rm \AA^{-1}$ from $\qvec = \kpoint$,
competition between the intraband and interband excitations take place
as shown in Figs.~\ref{Fig5}(a), (c), (d), and (f). According to the
analytical formula~\cite{sasaki12}, the intraband contribution to the
frequency shift is proportional to $-\sin^{-1}|2\efermi/vq|$ by
assuming $\omega_0\ll vq$, where $v$ is the slope of the linear energy
dispersion of graphene which is $\sim 6.46\ \rm eV\AA$. The phonon
linewidth is increasing linearly with $|\efermi|$ in the case of the
intraband excitation [Figs.~\ref{Fig5}(d) and (f)] because the
electron-phonon scattering rate is proportional to the carrier
concentration. The asymmetry at positive and negative $\efermi$ is
related to electron-hole band asymmetry considered in the tight
binding calculation.

\begin{figure*}[t]
\includegraphics[width=16cm]{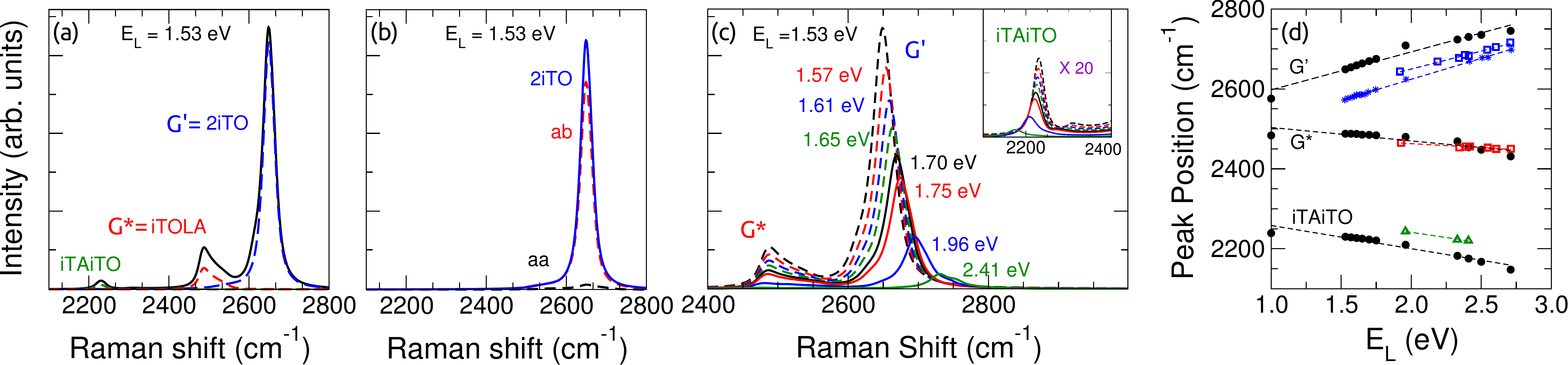}
\caption{\label{Fig7}(Color online) (a) The calculated results of the
  second-order Raman intensity for a laser energy 1.53 eV showing
  three significant peaks identified with the iTA+iTO band
  ($\sim2200\unitcm$), iTO+LA or $\Gstar$ band ($\sim2500\unitcm$),
  and 2iTO or 2D or $\Gprime$ band ($\sim2800\unitcm$). (b) The
  constituents of the $\Gprime$ band contribution from (a): $ab=
  (A^{\rm eh}+A^{\rm he})$ (blue dashed line) and $aa= (A^{\rm
    ee}+A^{\rm hh})$ (red dashed line). (c) The calculated results of
  the $\elaser$ dependence of the $\Gprime$, $\Gstar$, and $\iTAiTO$
  bands (inset) for $1.53 \unitev\le \elaser\le 2.41\unitev$. (d) The
  $\Gprime$, $\Gstar$, and $\iTAiTO$ bands peak position as a function
  of $\elaser$. Black dots are the calculated results (this work),
  blue and red open squares are from Ref.~\onlinecite{mafra07}, blue
  asterisks are from Ref.~\onlinecite{berciaud13}, and green triangles
  are from Ref.~\onlinecite{rao11}.}
\end{figure*}

After considering the KA effect on the $\qvec \neq 0$ phonon, in
Fig~\ref{Fig7} we show the calculated Raman spectra from
Eq.~\eqref{eq:twoph}. Figure~\ref{Fig7}(a) shows three bands,
respectively, assigned as the $\Gprime$ $\sim2700\unitcm$, $\Gstar$
$\sim2500\unitcm$, and $\iTAiTO$ $\sim2240\unitcm$ for
$\elaser=1.53\unitev$. We confirm the origin of the $\Gprime$ bands
from the overtone of the iTO (2iTO) modes while the $\Gstar$ bands
come from a combination of iTO and LA modes. The major contributions
to the $\Gprime$ intensity come from the $A^{\rm eh}$ and $A^{\rm he}$
terms as shown by $ab= (A^{\rm eh}+A^{\rm he})$ in
Fig.~\ref{Fig7}(b). This confirms the previous calculation that the
$A^{\rm ee}$ and $A^{\rm hh}$ terms are negligible [$aa= (A^{\rm
    ee}+A^{\rm hh})$ in Fig.~\ref{Fig7}(b)] because of the quantum
interference effect during the $\kvec$ integration~\cite{venezuela11}.

\begin{figure*}[t]
\includegraphics[width=16cm]{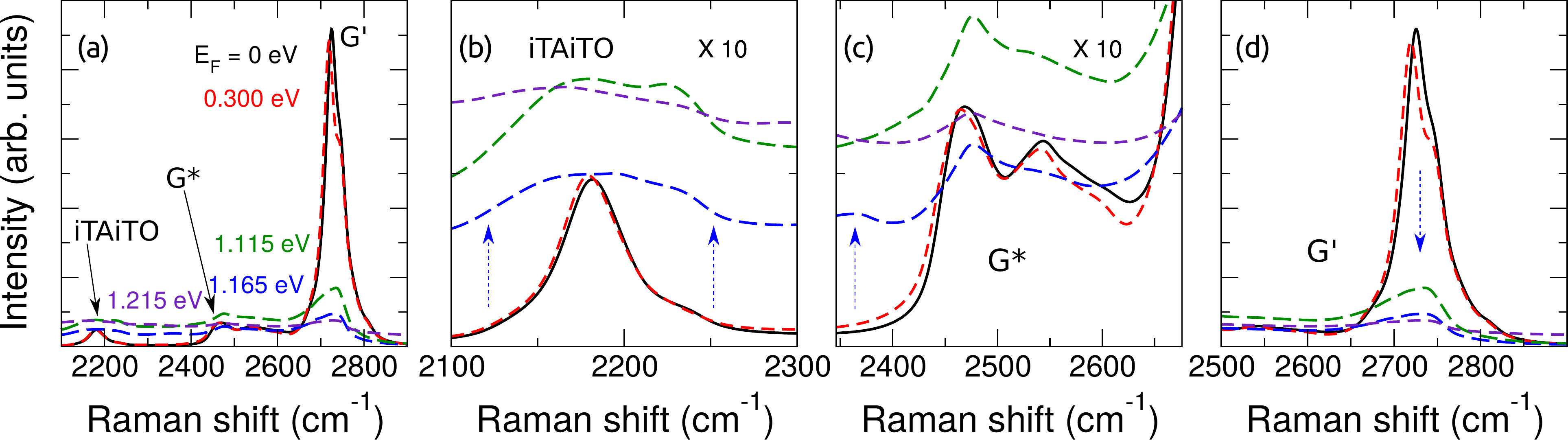}
\caption{\label{Fig8}(Color online) Raman spectra of (a) all
  second-order Raman bands (b) $\iTAiTO$, (c) $\Gstar$, and (d) $\Gprime$
  bands for $\efermi = 0$ (black solid lines), $\efermi = 0.3\unitev$
  (red dashed lines), $\efermi = 1.115\unitev$ (green dashed lines),
  $\efermi = 1.165\unitev$ (blue dashed lines), $\efermi =
  1.215\unitev$ (violet dashed lines). $\elaser$ used in this
  calculation is $2.33\unitev$.}
\end{figure*}

Figure~\ref{Fig7}(c) shows the second-order Raman intensities for
$1.53 \unitev\le \elaser\le 2.41\unitev$. The intensities of all these
Raman bands are inversely proportional to $\elaser$ because of the
increase of the electron-phonon scattering rate $\gamma$ as a function
of $\elaser$~\cite{venezuela11, liu15}. Assuming that each band can be
represented by a single peak, the $\Gprime$, $\Gstar$, and $\iTAiTO$
peak dispersions as a function of $\elaser$ are shown in
Fig.~\ref{Fig7}(d). The $\Gprime$ band shows a positive slope as a
function of $\efermi$, i.e., $95\unitcm/\unitev$ in this work,
$90\unitcm/\unitev$ in Ref.~\onlinecite{mafra07}, and
$104\unitcm/\unitev$ in Ref.~\onlinecite{berciaud13}. Meanwhile the
$\Gstar$ band shows a negative slope, i.e., $-33\unitcm/\unitev$ in
this work and $-33\unitcm/\unitev$ in Ref.~\onlinecite{mafra07} and
the $\iTAiTO$ band slope is $-58\unitcm/\unitev$ in this work,
$-56\unitcm/\unitev$ in Ref.~\onlinecite{rao11}, and
$-50\unitcm/\unitev$ in Ref.~\onlinecite{mafra12} [not shown in
  Fig.~\ref{Fig7}(d)]. Good agreement between theory and experiment in
the slope of the $\elaser$ dispersion indicates the reliability of our
phonon dispersion used in the calculation. However, discrepancies with
the experiments of about $50\unitcm$ in the $\Gprime$ and the
$\iTAiTO$ bands for a given $\elaser$ show that the calculated
electronic energy dispersion underestimates the experimenal
results. This can be seen insofar as the $\Gprime$ and $\iTAiTO$ peaks
at $\elaser=1.5\unitev$ in theory give relatively the same value for
$\elaser = 2.0\unitev$ in the experiment, thus the present electronic
energy dispersion near $\elaser$ underestimate the real value by
$\sim0.5\unitev$. This might be because we neglect the many body
effects in the band calculations.  Nevertheless, the overal agreement
is sufficient for us to proceed and consider the $\efermi$ dependence
of the Raman intensity for a particular $\elaser$.

\begin{figure*}[t]
\includegraphics[width=16cm]{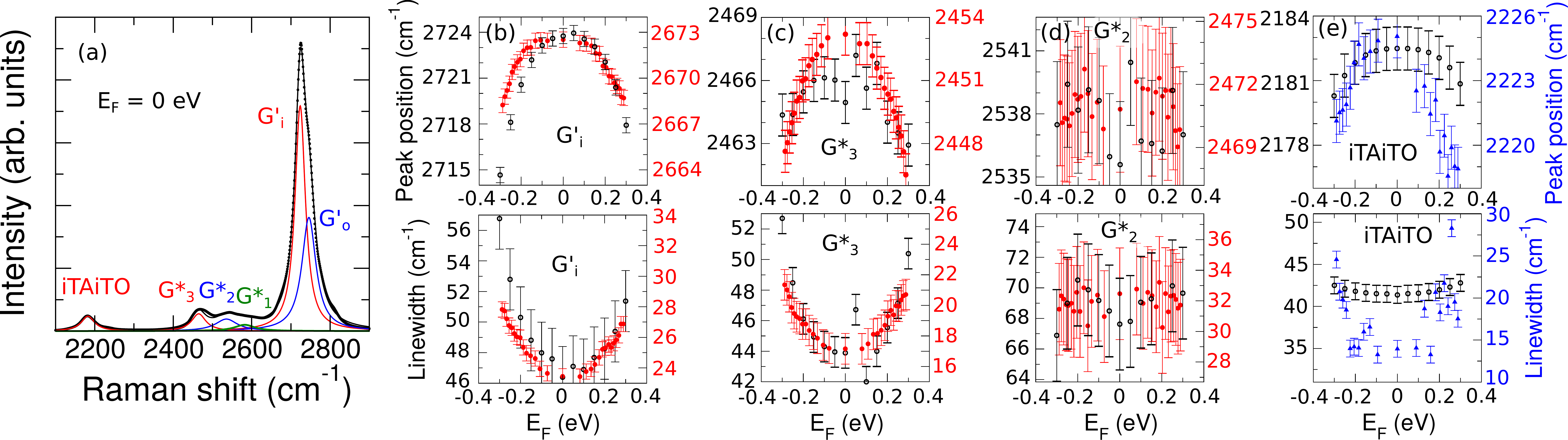}
\caption{\label{Fig9}(Color online) (a) Fitting of the second order
  Raman spectra obtained in Fig.~\ref{Fig8}(a) at $\efermi=0\unitev$
  and $\elaser=2.33\unitev$. The dotted line is the calculated Raman
  intensity fitted by six Lorentzians labelled by $\Gprime_o$ (blue),
  $\Gprime_i$ (red), $\Gstar_1$ (green), $\Gstar_2$ (blue), $\Gstar_3$
  (red), and $\iTAiTO$ bands.  We show the peak shift and the spectral
  linewidth as a function of $\efermi$ for (b) the $\Gprime_i$, (c)
  $\Gstar_3$, (d) $\Gstar_2$, and (e) $\iTAiTO$ bands. Black open
  circles are the results in this work, red closed circles are
  experimental results from Ref.~\onlinecite{araujo12}, and blue
  triangles are experimental results from Ref.~\onlinecite{mafra12}. }
\end{figure*}

Figure~\ref{Fig8} shows the evolution of the second order Raman
spectra for several values of $\efermi$: $\efermi = 0$ (black solid
lines), $\efermi = 0.3\unitev$ (red dashed lines), $\efermi =
1.115\unitev$ $(2\efermi=\elaser-0.1\unitev)$ (green dashed lines),
$\efermi = 1.165\unitev$ $(2\efermi=\elaser)$ (blue dashed lines),
$\efermi = 1.215\unitev$ $(2\efermi=\elaser+0.1\unitev)$ (violet
dashed lines). We use the same $\elaser=2.33\unitev$ as
Araujo~\emph{et al}~\cite{araujo12}. Figures~\ref{Fig8}(a) and (d)
show that the $\Gprime$ Raman spectrum is redshifted, decreases its
peak intensity [indicated by a blue arrow in Fig.~\ref{Fig8}(d)] and
broadens as $\efermi$ increases. These results are consistent with the
experiment~\cite{chen11, liu13}. However, the Raman intensity of the
$\Gstar$ and $\iTAiTO$ bands dramatically increases especially when
$2\efermi \approx \elaser$ [indicated by the blue arrows in
  Figs.~\ref{Fig8}(b) and (c)]. We find that the increase of the
intensity comes from the quantum interference effect and not from the
KA effect. Although the KA effect broadens the second-order Raman
spectra at a finite $\efermi$, however the broadening order should not
be $\sim100\unitcm$. Figures~\ref{Fig8}(a), (b), and (c) show that the
spectra where combination of the phonon modes are located
($\sim2000-2600\unitcm$) dramatically broadens like a continous
spectrum. The broadening should come from the increased intensity of
many different $\qvec$ phonon modes that are suppressed at $\efermi=0$
during integration on $\kvec$ but now appears at
$2\efermi\approx\elaser$ thanks to the suppression of the destructive
interference. These results give clues about how the Raman phase
governed by the electron-phonon matrix elements distinguishes between
the $\efermi$ dependence of the Raman intensity of the combination
modes from that of the overtone modes as is also observed in
experiments~\cite{liu13,bruna14}. Although $\efermi\approx 1\unitev$
is too high for experiments on graphene, one can reduce the $\elaser$
to become $ \approx 1\unitev$ to satisfy the condition of $2\efermi
\approx \elaser$ to get proper conditions for observing the quantum
interference effect.

The calculated Raman spectra at $\efermi = 0\unitev$ and
$\elaser=2.33\unitev$ in Fig.~\ref{Fig8}(c) and (d) clearly show that
the $\Gstar$ and $\Gprime$ bands cannot be fit by a single Lorentian
for each band. Figure~\ref{Fig9}(a) shows the Lorentzian fitting
results on the second order Raman spectra for $\efermi=0$. The dotted
line is the calculated Raman intensity fitted by six Lorentzians.  We
fit the $\Gprime$ bands with two Lorentzians labelled by $\Gprime_o$
(blue) and $\Gprime_i$ (red) which reffer to $\Gprime$ bands from
outer ($\qvec$ in KM direction) and inner ($\qvec$ in $\kpoint\Gamma$
direction) scattering processes, respectively~\cite{venezuela11,
  berciaud13}. Three Lorentzians are needed to fit the $\Gstar$ band,
labelled by $\Gstar_1$ (green), $\Gstar_2$ (blue), and $\Gstar_3$
(red). Finally one Lorentzian is used to fit the $\iTAiTO$ band.

After Lorentzian fitting, we compare both the peak shift and the
spectral linewidth as a function of $\efermi$ as shown in
Figs.~\ref{Fig9}(b)-(e). We do not show the $\Gprime_o$ and $\Gstar_1$
for simplicity because there is no experimental data available for
comparison. The calculated results in Fig.~\ref{Fig9}(b)-(e) cannot
fit the experimental value of both the peak position and the linewidth
due to the underestimation of the electronic energy dispersion as
previously discussed in the $\elaser$ dependence of the second-order
Raman spectra~[see Fig.~\ref{Fig7}(d)].  But we can discuss the change
of both quantities as a function of $\efermi$, where the KA effect
takes place. In Figs.~\ref{Fig9}(b)-(e), both the spectral peak
position and the linewidth as a function of $\efermi$ are plotted in
the same range, comparing the theory and experiments. Reasonable
agreements between experiments and theory are achieved. The three
major peaks, i.e., the $\Gprime_i$, $\Gstar_3$, and $\iTAiTO$ bands
show ``$\Lambda$'' (``V'') shapes of the Raman peak shift (spectral
linewidth) as a function of $\efermi$. These behaviors exist because
of the intraband electron-hole excitation renormalization of phonons
as shown in Fig.~\ref{Fig5}. The $\Gstar_2$ band in Fig.~\ref{Fig9}(d)
is relatively dispersionless in $\efermi$ because it is located in the
shoulder of the $\Gprime$ band where 2iTO $\qvec=K$ exists. Therefore,
for these bands, the competition between interband and intraband
electron-hole excitations are expected. The calculated results
overestimate the experimental spectral linewidths of all bands, which
is related to the choice of $\Delta\kvec$ in the $\kvec$ integration.
We can tackle this issue by reducing the value of $\Delta \kvec$ by
$\Delta \kvec/n$; however, the computational burden becomes $e^n$
times larger.

\section{Conclusion}
\label{sec:conclusion}
In conclusion, we calculated the first- and second-order Raman spectra
as a function of $\efermi$. The opposite effect of the Kohn anomaly
that is found experimentally between the first- and the second-order
Raman spectra occurs because the KA effect on the first-order Raman
spectra is dominated by the renormalization of phonons by the
interband electron-hole excitation, while in the second-order Raman
spectra, the intraband electron-hole excitation dominates over the KA
effect. We also discussed the quantum interference effect observed in
the change of the Raman intensity as a function of $\efermi$. Both the
first- and the second-order Raman spectra exhibit an impact of the
quantum interference effect, especially when
$2|\efermi|\approx\elaser$. Present calculated results found that not
only is the resonance condition important, but also the explicit
consideration of the electron-phonon matrix elements are essential to
determine the $\efermi$ dependence of the Raman spectral lineshape.

\section*{Acknowledgments}
E.H. is supported by a MEXT scholarship. A.R.T.N. acknowledges the
Interdepartmental Doctoral Degree Program for Material Science Leaders at
Tohoku University for providing a financial support.  R.S.
acknowledges MEXT Grants No. 25286005 and No. 225107004.
M.S.D. acknowledges NSF-DMR Grant No. 15-07806.

% \bibliographystyle{aip}
% \bibliography{stda2000.bib,stdb2000.bib}

\end{document}